%% file: arxiv_main.tex
\documentclass[notitlepage]{article}
\usepackage{natbib}
\usepackage{booktabs}
\usepackage{csquotes}
\usepackage{tikz}
\usepackage{hyperref}
\usepackage{amsmath,amssymb,mathtools}
\usepackage{amsthm}
\newtheorem{theorem}{Theorem}[section]

\newtheorem{proposition}[theorem]{Proposition}
\newtheorem{assumption}[theorem]{Assumption}
\newtheorem{lemma}[theorem]{Lemma}
\newtheorem{corollary}[theorem]{Corollary}
\newtheorem{conjecture}[theorem]{Conjecture}
\newtheorem{definition}[theorem]{Definition}
\newtheorem{observation}[theorem]{Observation}

\newcommand{\Wel}{\operatorname{Wel}}
\newcommand{\OPT}{\operatorname{OPT}}
\newcommand{\Rev}{\operatorname{Rev}}
\newcommand{\R}{\mathbb{R}}
\newcommand{\argmax}{\operatorname{arg\,max}}

\setcitestyle{authoryear}
\begin{document}
\title{Certification Design for a Competitive Market}
\author{Andreas Haupt \and Nicole Immorlica \and Brendan Lucier\thanks{We thank Robert Maddox, Yanika Meyer-Oldenburg and a seminar audience at Harvard.}}
\maketitle
\begin{abstract}
Motivated by applications such as voluntary carbon markets and educational testing, 
we consider a market for goods with varying but hidden levels of quality in the presence of a third-party certifier. The certifier can provide informative signals about the quality of products, and can charge for this service. Sellers choose both the quality of the product they produce and a certification. Prices are then determined in a competitive market. Under a single-crossing condition, we show that the levels of certification chosen by producers are uniquely determined at equilibrium. We then show how to reduce a revenue-maximizing certifier's problem to a monopolistic pricing problem with non-linear valuations, and design an FPTAS for computing the optimal slate of certificates and their prices.  In general, both the welfare-optimal and revenue-optimal slate of certificates can be arbitrarily large.
\end{abstract}
\input{intro.tex}
\input{litrev.tex}
\input{model.tex}
\input{equil.tex}

\input{revenue.tex}

\input{comparison.tex}

\bibliographystyle{ACM-Reference-Format}
\bibliography{refs}
\end{document}

%% file: intro.tex
\section{Introduction}
Many markets sell goods that have different features but appear identical to consumers.  Examples include the market for carbon credits, the market for contract workers such as electricians or plumbers, or even some markets for physical goods like milk and eggs.  In each of these markets, the goods have hidden, hard-to-verify properties that distinguish them.  Nature-based carbon credits, for example, are sourced from a variety of forests with different protection and longevity properties.  Contract workers have differing skill levels and amount of knowledge.  And physical goods are produced in a variety of circumstances, some more ethical and sustainable than others.

In such settings, sellers seek to distinguish their products through costly certification.  In the example of carbon credits, large certifiers  issue certification standards for carbon emissions. Contract workers can enroll in certification programs or take exams to document their skill level.  And farms can reach out to third-party organizations that certify the conditions, such as free-range or vegetarian-fed, under which their food is produced. 

In each of these cases, a downstream market allocates goods based on certification. Carbon exchanges allow for the trade of carbon certificates, and platforms for local services allow the trade of electrical or plumbing services. These markets have price formation that is independent of, and not controllable by, the certifier. Therefore, both welfare- and revenue-maximizing certifiers will need to reason about how the certification options they provide affect trade, and how this in turn influences the demand for certification.

What impact does the presence of certification have on the downstream market? We answer this question in the context of a competitive market with production. In our model there is a mass of heterogeneous producers each creating a single unit of a vertically-differentiated product.  Producers can select the quality level of their products.  The corresponding production cost is determined by the producer's type and is increasing in quality. A mass of unit-demand consumers purchases these goods. Consumers also have varying types which determine the strength of their preference for quality. Types are totally ordered and satisfy a single-crossing condition: higher-type producers are relatively more efficient at producing higher-quality goods, which higher-type consumers have relatively stronger preference for.

Quality cannot be verified directly; instead, a third-party certifier offers a menu of certifications, each with its own requirements and certification price.  Producers can purchase certifications of their products' qualities from this menu.  If their product surpasses the quality level of the certificate, it will be marketed as such. The certified goods, together with indistinguishable uncertified ones, are then sold in a competitive market.

As in the Economics literature on certification, we assume that market participants are able form correct beliefs about the quality implications of a certificate \cite{milgrom1981good,grossman1981informational,lizzeri1999information,demarzo2019test}. A big strand of empirical work considers the question of whether certifications are correctly interpreted in various markets (\cite{wimmer2003empirical} for race horses, \cite{tadelis2015information} on used-car auctions, \cite{ramanarayanan2012reputations} for dialysis screening centers, \cite{luca2016reviews} for restaurant reviews, \cite{elfenbein2015market} for seller ratings on Ebay, \cite{conte2010explaining} for voluntary carbon credits). In our model, certificates are based on verifiable certification requirements, and at perfect Bayesian equilibrium all market participants hold rational beliefs about the distribution of quality implied by any given certificate. We note that such rational beliefs need not be aligned with official certificate descriptions.  For example, in voluntary carbon markets, several empirical contributions point out that carbon certificate descriptions might not correspond to counterfactual mitigation outcomes, compare \cite{west2020overstated,west2023action,guizar2022global,unearthed,guardian,guardian2}.

The presence of certificates clearly impacts the market equilibrium because it influences consumers' beliefs and hence willingness-to-pay.  Our first result shows that, for any menu of certificates offered by the certifier, the equilibrium choice of certificates and the allocation outcome of the downstream market equilibrium is unique.  Furthermore better producers (that is, those who can produce higher quality at lower cost) always purchase higher (more restrictive) levels of certification.  This implies that the producer-consumer matching in equilibrium is assortative.  Better producers sell to consumers with higher value for quality, and at higher quality levels.

This analysis immediately implies that the welfare-optimal menu offers every certificate at cost, and suggests a dynamic program for finding the welfare-optimal menu subject to a cardinality restriction on the number of certification levels that can be offered. But in many markets, certification is provided by third-party firms with profit-maximizing incentives.  Our main result shows how to approximately construct the revenue-optimal menu in polynomial time. Specifically, we provide an FPTAS: we show how to compute a menu whose revenue for the certifier is an additive $\epsilon$ less than the optimal revenue in time polynomial in $1/\epsilon$.  Our algorithm can optionally take as input a cardinality constraint $k$ on the menu size (i.e., a maximum number of certification levels), in which case the revenue benchmark is the optimal slate of at most $k$ certificates and their prices.

Our proof contains technical insights that may be of independent interest.  Namely, we reduce the certifier's problem to one of a seller who wishes to sell a divisible good, facing buyers whose valuations are non-linear and potentially non-monotone in quantity.  The seller corresponds to the certifier, and the buyers correspond to producer-consumer pairs.  This projection of producers and consumers into a single economic agent is enabled by the fact that the matching in equilibrium is unique and assortative. The buyer valuations in this reduced problem are concave but not necessarily monotone as a function of quantity, meaning that for different buyer types the valuation may reach its maximum at different quantities.  The valuations do satisfy a single-crossing property, which implies that types are totally ordered and higher-type buyers have pointwise higher valuations and have weakly higher preferred quantities.  Under these conditions, we show that the revenue-optimal menu may be non-linear but will exhibit prices that are monotone in quantity.  Monotonicity of prices enables the use of dynamic programming to construct an approximately optimal menu, though some care must be taken when discretizing the space of quantities and prices to ensure that buyer preferences do not change substantially.  This non-linear pricing problem generalizes prior work that specifically studied the case of quadratic demand~\cite{bergemann2012multi,bergemann2012mechanism}, and may be of independent interest.

The menu of certifications offered by a revenue-maximizing third-party certifier can distort welfare and result in inefficient levels of trade.  That said, we show that a certifying agent entering into the market can never lead to a loss of welfare, no matter what menu of certification options they make available.  More specifically, we can imagine that a collection of (perhaps costly) certification options are  available to the market as a baseline, perhaps provided by a public or tightly regulated agency.  If a third-party certifier then enters the market and makes available any additional certification options into the market, at any price, we show that the total utility of all market participants (even excluding the new entrant) can only increase.  A policy implication is that a regulatory body concerned about inefficient certification arising from profit-seeking motives need not prevent certifiers from entering the market.  Rather, it suffices to make sure that there exists some set of certifications available, either through third-party providers or public options, that enable an acceptable level of welfare from trade.

%% file: litrev.tex
\subsection{Related Literature}\label{subsec:litrev}
This work contributes to the literature on certification. The early results \cite{milgrom1981good,grossman1981informational} produce unraveling type results: In equilibrium, the quality of a good is fully revealed. The main intuition for these results in models of certification is that certifying non-informatively will be interpreted by the market as a sign of bad quality---adverse selection is extreme. In our model of a revenue-maximizing certifier, there might be other reasons for certification less informatively, as the price in the competitive market may depend on not only an individual seller's quality, but all the sellers in the market. Later contributions \cite{lizzeri1999information,demarzo2019test,acharya2011endogenous} allow for unsuccessful certifications, restricting the stark result obtained in \cite{milgrom1981good,grossman1981informational}.

More broadly, our work is related to mechanism and information design in the presence of an exogenously given game played after the design.  We recommend \cite{bergemann2019information} for a general overview of information design.  
Our work is especially related to the design of information provided to buyers and/or sellers of a good. \cite{bergemann2017first} considers the design of information for a first-price auction, where a third party can reveal a signal correlated with a buyer's valuations, and fully characterizes the achievable revenue and payoffs.  \cite{alijani2022limits} extends the analysis to a scenario with multiple buyers. In a general mechanism design framework,~\cite{CandoganS21} develop an optimal disclosure policy for action recommendations in a game with private types and a hidden state. \cite{dworczak2020mechanism} designs mechanisms in a setting where players participate in a finite Bayesian game after participating in the mechanism, so that game outcomes are impacted by information revealed over the course of the mechanism. He finds a cutoff structure of optimal mechanisms in the first stage.

For-profit certification relates to the sale of hard information, which has been studied in the context of competitive markets.  \cite{ali2022sell} consider a seller holding a good of uncertain quality.  The seller can purchase a quality-correlated signal from a revenue-maximizing intermediary before bringing the good to market.  In general the resulting equilibria are not unique and can vary substantially, but by employing noisy signals the intermediary can robustly guarantee high revenue.  Our model differs in that any certification options are made available to the entire market and product quality is endogenous, so the certifying agent can impact welfare.
More generally, our work relates to the problem of how to sell payoff-relevant hard information to a potential buyer.  \cite{bergemann2018design} solve for the revenue-maximizing mechanism in binary environments, and~\cite{bergemann2022selling} establish when full disclosure is approximately optimal under more general spaces of actions.  In contrast, our certifying agent is selling a signal that is valuable in that it conveys information to other participants in a subsequent game.

In our mathematical analysis, we reduce the certifier's problem to a pricing problem with a non-linear valuation. The non-linear pricing literature following \cite{mussa1978monopoly} (see also treatment in \cite{dewatripont2005contract} and \cite[Chapter 2.3]{borgers2015introduction}) studies a non-linear concave valuation with a quadratic cost. The functional form assumptions in these papers allow to characterize the optimal mechanism in closed form. Typically, the optimal menu of offered goods is a continuum, in contrast to the linear screening problem first studied in the influential \cite{myerson1981optimal}. Our analysis will show that also the class of models we consider may feature infinite menus.

In our polynomial-time approximation scheme, we use an approximation of a non-linear pricing model. The papers \cite{bergemann2012multi,bergemann2012mechanism} consider approximation of non-linear single- and multi-dimensional pricing environments with finite menus (in the papers called \enquote{finite information}). The papers make functional form assumptions similar to the ones in \cite{mussa1978monopoly}, and derive rates of approximation by finite menus in these finite menus. The present paper allows for a general class of utility functions that satisfy a single-crossing condition, and, in addition to showing approximation by finite menus, shows that the finite-sized menu can be computed efficiently.

Finally, our main assumption guaranteeing uniqueness of our equilibrium is a single-crossing condition. Single-crossing conditions are important in several domains, among them the interdependent private values auctions (\cite{milgrom1982theory}) and social choice and voting (\cite{saporiti2006single}).  A recent line of work in algorithmic mechanism design has employed single-crossing conditions to enable approximately optimal designs in interdependent value settings~\cite{roughgarden2013optimal,chawla2014approximate}.  Closest to the present paper, another implication of single-crossing is adverse selection in markets \cite{mirrlees1971exploration,spence1974competitive}.

\subsection{Outline}
The rest of this article is structured as follows.  We formalize our model in \autoref{sec:model}. In \autoref{sec:eq} we analyze the structure of equilibria given the certifier's offerings. 
Section \ref{sec:rev} contains our main results, a reduction of revenue-maximizing certification to a non-linear pricing problem, the FPTAS for its computation.  Section \ref{sec:comparison} contains our results on welfare maximization and explores the welfare implications of third-party certification.

%% file: model.tex
\section{Model}\label{sec:model}

\paragraph{Market}
We consider a continuum market between producers and consumers.  Producers are unit-supply and parameterized by types $\psi \in \R_+$ with measure $G$. Consumers are unit-demand and parameterized by types $\phi \in \R_+$ with measure $F$.  The type measures $F$ and $G$ are atomless and continuous with compact support.

Goods can be produced at different levels of quality, denoted $q \in [0,1]$.  Goods of higher quality are more valuable to consumers but more costly to produce.  We write $g(q; \psi)$ for the cost incurred by a producer of type $\psi$ when producing a good of quality $q$.  We assume $g$ is weakly convex and non-decreasing in $q$ for every $\psi$ and normalized so that $g(0; \psi) = 0$.
We also write $f(q; \phi)$ for the value enjoyed by a consumer of type $\phi$ for a good of quality $q$, where $f$ is weakly concave and non-decreasing in $q$ for every $\phi$ and normalized so that $f(0; \phi) = 0$.  We will scale valuations so that $f(q; \phi) \leq 1$ for all $q$ and $\phi$, which is without loss for bounded values.

We will assume that costs and valuations satisfy single-crossing with respect to the producer and consumer types, respectively.  Roughly speaking, this means that producers (consumers) of higher types have lower marginal cost (higher marginal value) for producing higher-quality goods.  More formally, for all $\phi_1 < \phi_2$ and $q_1 < q_2$, we have
\[ f(q_2; \phi_2) - f(q_1; \phi_2) > f(q_2; \phi_1) - f(q_1; \phi_1). \]
Likewise, for all $\psi_1 < \psi_2$ and $q_1 < q_2$, we have
\[ g(q_2; \psi_2) - g(q_1; \psi_2) < g(q_2; \psi_1) - g(q_1; \psi_1). \]

Transfers between producers and consumers are permitted.  We assume that producers and consumers are risk-neutral and have quasi-linear preferences with respect to money.  That is, if consumer $\phi$ purchases a product of quality $q$ from producer $\psi$ at a price of $p$, then the consumer enjoys utility
\[ u_C((q,p); \phi) = f(q; \phi) - p \]
and the producer's utility is
\[ u_P((q,p); \psi) = p - g(q; \phi). \]

\paragraph{Certification}
Crucially, producers and consumers cannot contract on quality.  This means that a producer cannot credibly commit to the quality of the good they produce, and a consumer cannot verify quality at the point of trade.  But there is a third-party certifier who is able to determine the quality of a producer's good.  This verification is costly to the certifier, with cost $c \geq 0$.

After verifying the quality of a good, the certifier is able to assign to that good a signal (or certificate) $\sigma \in \Sigma$, where $\Sigma$ is an arbitrary space of potential certificates.  This certificate is visible to all producers and consumers.  The certifier is permitted to collect payments from producers for this service, and these transfers can depend arbitrarily on the certificate produced.

The certifier can commit to a certification menu $M$, which is a collection of certificate / transfer pairs $(\sigma, t(\sigma))$ along with quality requirements for each certificate $\sigma$.  Following the literature on information design and persuasion, we note that it is without loss of generality to associate each certificate signal $\sigma$ with the set of quality levels that are eligible for that certificate.  We will therefore assume without loss of generality that $\Sigma = 2^{[0,1]}$, the collection of all subsets of $[0,1]$, and each $\sigma$ is a subset of $[0,1]$.  The interpretation is that a producer can select an item from this menu, in which case she pays the certifier price (or transfer) $t(\sigma)$, the certifier verifies the good's quality $q$, and as long as $q \in \sigma$ the producer will receive certification $\sigma$.  We will assume for technical convenience that each $\sigma$ in the certifier's menu contains a minimum element, meaning that $\inf \sigma \in \sigma$.\footnote{This excludes certificates of the form ``the quality $q$ is strictly greater than $1/2$," as opposed to ``...at least $1/2$."  Certificates of the former type are inconvenient because there is no single least-cost choice of quality that satisfies the certification requirement, and hence no utility-maximizing choice of quality for the producer.  We could handle such certificates in a straightforward but tedious way by relaxing our equilibrium notion and assuming that each producer selects an $\epsilon$-approximately utility maximizing choice of quality for some arbitrarily small $\epsilon$.  For the remainder of the paper we will ignore such technical issues and simply assume that each $\sigma$ includes a minimum element.}

We will write $\sigma_0 = [0,1]$ for the trivial signal that conveys no additional information about quality.  A good with certificate $\sigma_0$ is functionally equivalent to a good that has not been certified.  It will be notationally convenient to assume that the certifier always offers signal $\sigma_0$ at cost $0$.  This allows us to think of every good as being certified, albeit possibly at the trivial level.  If a producer attempts to purchase a certificate but does not satisfy that certificate's requirements, they will instead be assigned certificate $\sigma_0$; i.e., the certifier will not certify the good.

\paragraph{The Competitive Market} 
All certification is assumed to occur simultaneously, and in advance of any trading between producers and consumers.  Given the menu $M$ of certificates and prices offered by the certifier, the producers' (production and certification) strategy is a mapping from producer type $\psi$ to a choice of quality level $q$ and certification $\sigma$.  We will denote such a strategy $\Gamma: \psi \mapsto (q,\sigma)$, and restrict attention to measurable functions $\Gamma$.  In a slight abuse of notation, we will also use $\Gamma$ to denote the measure over pairs $(q,\sigma)$ of products with corresponding quality and certification that result when producers apply strategy $\Gamma$.

Goods that are assigned the same certificate are indistinguishable by the consumers.  After all certification is complete, each producer has a single unit of a good marked with a certification $\sigma$. For any given certification $\sigma$, write $\Gamma_\sigma$ for the marginal distribution over quality $q$ of $\Gamma$ restricted to certificate $\sigma$.  That is, fixing the choices of the producers, $\Gamma_\sigma$ is the distribution of levels of quality for a product with certification $\sigma$.  
Then the value of a consumer of type $\phi$ for a good with certificate $\sigma$ can be evaluated as
\[ f(\sigma; \phi) = E_{q \sim \Gamma_\sigma}[f(q; \phi)]. \]
That is, each consumer rationally evaluates the expected quality of each product given its certification level and the choices of the producers.

Since goods with the same certification are indistinguishable to consumers, we can view the competitive market between producers and consumers as a market for certificates $\sigma$.  A Walrasian (or Competitive) equilibrium of the resulting market is an allocation $x(\phi)$ of a certificates to each consumer $\phi$, along with a price $p_\sigma$ for each certificate, such that:
\begin{itemize}
    \item Demand satisfaction: every consumer purchases her most-preferred good.  That is, for every consumer type $\phi$, $f(x(\phi); \phi) - p_{x(\phi)} \geq f(\sigma; \phi) - p_\sigma$ for every $\sigma \in \Sigma$.
    \item Market Clearing: every good with a positive price is sold.  That is, for all $\sigma$, the measure of consumers $\phi$ such that $x(\phi) = \sigma$ is at most the measure of producers $\psi$ who select level of certification $\sigma$.  If $p_\sigma > 0$ then these measures are equal.
\end{itemize}
Since buyers (consumers) are unit-demand and hence their preferences satisfy the gross substitutes condition, a Walrasian equilibrium is guaranteed to exist~\cite{gul2000english}.  We will therefore assume that trade occurs between consumers and producers at competitive equilibrium prices given the choices made by the producers.

\paragraph{Timeline} To summarize, the timing of the market with certification is as follows:
\begin{enumerate}
    \item The certifier commits to a menu of certificates with corresponding prices.
    \item Each producer $\psi$ simultaneously and privately chooses whether to produce a good, and if so at what level of quality.
    \item Each producer that chose to produce decides whether to certify their product, and at which certificate.  These decisions are made simultaneously for all producers.  
    \item The certifier verifies the products of producers who choose to certify and assigns certificates.  Any producer who does not successfully certify receives certificate $\sigma_0$.
    \item Producers and consumers trade goods in a competitive market.  I.e., trade occurs at market-clearing prices for the chosen levels of certification.
\end{enumerate}

Since Walrasian equilibria are not unique in general, one might wonder if the outcome described in the final step is well-defined.  We will show in the next section that the competitive market equilibrium described the final step exists and its resulting allocation is unique for any certificate menu chosen by the certifier and any choice of certification levels chosen by the producers.  

%% file: equil.tex
\section{Equilibrium Characterization}\label{sec:eq}
In this section we describe the market outcome that will occur for any given menu of certificates offered by the certifier.  We show that it is without loss of generality for the certifier to restrict to offering threshold certificates that guarantee that a product is at least a certain level of quality. We characterize the unique Walrasian market equilibrium allocation that results from any assignment of such certificates to producers.  We then use that characterization to solve for each producer's utility-maximizing choice of certificate from the certifier's menu, which will also be unique.

\subsection{Certifications as Minimum Quality Thresholds}

A first simple observation is that since producer costs are increasing in quality level, and since goods at different quality levels but with the same certification are indistinguishable to consumers (and hence must sell at the same price), a producer who is assigned certificate $\sigma$ will always choose to produce at the minimum quality level eligible for that certificate.

\begin{observation}\label{obs:min.quality}
Fix any certifier menu $M$ and any production and certification strategy of the producers.  Then for any producer $\psi$, selecting certificate $\sigma$ and producing at quality $q > \min\sigma$ is dominated by selecting certificate $\sigma$ and producting at quality $\min\sigma$.
\end{observation}

Given this observation, we know that for any menu $M$, any equilbrium strategy $\Gamma$ for the producers, and any certificate $\sigma$, the marginal distribution over quality $\Gamma_\sigma$ will be a point mass at $\min\sigma$.  In particular, any two certificates with the same minimum will induce the same equilibrium beliefs over quality and hence have indistinguishable value to all consumers.
It is therefore without loss of generality for the certifier to only offer certificates of the form $\sigma_q = [q,1]$; i.e., certificates that are differentiated only with respect to their minimum values.  If a producer selects certificate $\sigma_q$, then that producer's chosen quality level at equilibrium will necessarily be $q$.  Any given certification menu $M$ therefore reduces to a (possibly infinite) collection of quality levels in $[0,1]$ to certify.  

Motivated by this observation, 
we will assume for the remainder of the paper that all certificates are of the form $[q,1]$, and associate each $\sigma = [q,1]$ with its quality threshold $q$.
We can then think of a certification menu $M$ as a collection of pairs $\{(q_i, t_i)\}$, where $q_i$ is a quality threshold and $t_i$ is a corresponding price for certifying that quality is at least $q_i$.

\subsection{Uniqueness and Assortativeness of Competitive Market Allocations}

We now turn to an analysis of the competitive market outcome that will result given the strategies of the certifier and producers.  Recall that we can restrict attention to certificates of the form $\sigma_q = [q,1]$ and that any good with certificate $\sigma_q$ will have quality $q$ with probability $1$, so for the remainder of the section we will think of a market outcome as an allocation $x$ and prices $p$ of quality levels.  That is, $x(\phi) \in [0,1]$ for all consumers $\phi$, and for each $q \in [0,1]$ in menu $M$ there is an associated market price $p_q$. We emphasize that $x$ is a mapping from consumers to the certified goods they buy at market, whereas $\Gamma$ is a mapping from producers to the certificates that they choose from the certifier.

The following lemma shows that for any choice of certification menu $M$ and production and certification strategy $\Gamma$ of the producers, all competitive market equilibria in the resulting market have the same uniquely-determined allocation.  This allocation will be assortative, with higher-type consumers purchasing the higher-quality certificates.

\begin{lemma}
\label{lem:market.monotone.consumer}
Fix any certifier menu $M$ and any strategy $\Gamma$ of the producers. Then in every competitive market outcome $(x,p)$, the allocation $x$ satisfies $x(\phi_1) \leq x(\phi_2)$ and $p_{x(\phi_1)} \leq p_{x(\phi_2)}$ for all $\phi_1 \leq \phi_2$.
\end{lemma}
\begin{proof}
Since consumers are unit-demand and each producer has a single unit of good, a Walrasian equilibrium $(x,p)$ of the market is guaranteed to exist.  By the first welfare theorem, any such equilibrium must maximize the total welfare,
\[ \int_\phi f(x(\phi); \phi) dF(\phi). \]
Suppose there exist types $\phi_1 < \phi_2$ with $x(\phi_1) > x(\phi_2)$.  By the single-crossing condition, we have that
\[ f(x(\phi_1);\phi_2) - f(x(\phi_2);\phi_2) > f(x(\phi_1);\phi_1) - f(x(\phi_2);\phi_1) \]
and hence
\[ f(x(\phi_1);\phi_2) + f(x(\phi_2);\phi_1) > f(x(\phi_1);\phi_1) + f(x(\phi_2);\phi_2) \]
which contradicts the supposed welfare optimality of allocation $x$.

We now turn to prices.  Fix any consumer types $\phi_1 < \phi_2$, so in particular we know $x(\phi_1) \leq x(\phi_2)$, and suppose for contradiction that $p_{x(\phi_1)} > p_{x(\phi_2)}$.  By monotonicity of the value function $f$, we must have $f(x(\phi_1); \phi_1) \leq f(x(\phi_2); \phi_1)$.  But this then means $f(x(\phi_2); \phi_1) - p_{x(\phi_2)} > f(x(\phi_1); \phi_1) - p_{x(\phi_1)}$, which violates the competitive market condition that consumer type $\phi_1$ is choosing her most-preferred good.  We therefore conclude that $p_{x(\phi_1)} \leq p_{x(\phi_2)}$, as claimed.
\end{proof}

\subsection{Uniqueness and Assortativeness of Certificate Selection}

Given that market outcomes are well-defined, we next turn to the equilibrium choices of the producers when selecting quality levels and their corresponding certifications.  We again show that for any menu $M$ of certificates offered, the quality choices of producers are unique at equilibrium.  Moreover, higher-type producers will always select (weakly) higher certificates.  {In the Lemma~\ref{lem:market.monotone.consumer}, we saw that higher-type consumers also purchase (weakly) higher certificates.  As we discuss later, this means the matching in any equilibrium will be assortative and hence constrained-efficient given the available certificates.} 

Recall that a producer strategy $\Gamma$ is a mapping from producer type $\psi$ to a choice of certification and quality, which we know from will always coincide.  We will therefore write $\Gamma(\psi) = q$ to mean that producer $\psi$ produces at quality level $q$ and purchases certificate $\sigma_q$.  In particular, we must have $\Gamma(\psi) \in M$ for all $\psi$.

\begin{lemma}
\label{lem:market.monotone.producer}
Fix any certification menu $M$ offered by the certifier.  Then there is a unique equilibrium strategy $\Gamma$ for the producers, and $\Gamma(\psi)$ is weakly increasing in $\psi$.
\end{lemma}
\begin{proof}
Fix strategy $\Gamma$, which implies the measure of certificates chosen by the collection of producers.  Let $(x,p)$ denote a Walrasian equilibrium in the resulting competitive market, and recall that $x$ is uniquely determined.  

We first show that $\Gamma$ is weakly increasing is $\psi$.  Assume for contradiction that there exist $\psi_1 < \psi_2$ with $q_1 = \Gamma(\psi_1)$ and $q_2 = \Gamma(\psi_2)$ with $q_2 < q_1$.  Then by the single-crossing condition for producers, we have $g(q_1; \psi_1) - g(q_2; \psi_1) > g(q_1; \psi_2) - g(q_2; \psi_2)$.  But then, if we let $p_{q_1}$ and $p_{q_2}$ denote the Walrasian equilibrium prices of $q_1$ and $q_2$ given $\Gamma$, we have
\[ \left(p_{q_1} - g(q_1; \psi_1)\right) + \left(p_{q_2} - g(q_2; \psi_2)\right) < \left(p_{q_1} - g(q_1; \psi_2)\right) + \left(p_{q_2} - g(q_2; \psi_1)\right) \]
which means that either 
\[ p_{q_1} - g(q_1; \psi_1) < p_{q_2} - g(q_2; \psi_1) \]
or
\[ p_{q_2} - g(q_2; \psi_2) < p_{q_1} - g(q_1; \psi_2). \]
In other words, either producer $\psi_1$ or $\psi_2$ (or both) would strictly improve their utility by switching their choice of quality and certification.  As such a swap has measure zero and does not influence the competitive equilibrium, this would be an improving deviation for the producer(s), violating the assumption that $\Gamma$ is an equilibrium strategy for the producers. 


We have shown that $\Gamma$ is weakly increasing in $\psi$.  On the other hand, we know from Lemma~\ref{lem:market.monotone.consumer} that the market allocation of quality levels to consumers is weakly increasing in $\phi$.  This means that any equilibrium outcome of production and trade is equivalent to one in which consumers and producers are matched assortatively, with higher-type consumers trading with higher-type producers.
In other words, for any producer type $\psi$, there is a consumer type $\phi = \phi(\psi)$ such that $\psi$ always trades with $\phi(\psi)$.  Specifically, $\phi(\psi)$ is such that $F(\phi(\psi)) = G(\psi)$ (treating $F$ and $G$ as cumulative distribution functions).

Given this, we claim that $\Gamma(\psi)$, the certification selected by producer $\psi$ at equilibrium, will always be a certificate $q_i$ from the certifier's menu $M = \{(q_i, t_i)\}$ that maximizes $f(q_i; \phi(\psi)) - g(q_i; \psi) - t_i$.  To see why, suppose for contradiction that the producer instead chooses some other certificate $q'$ at price $t'$ such that $f(q'; \phi(\psi)) - g(q'; \psi) - t' < f(q_i; \phi(\psi)) - g(q_i; \psi) - t_i - \epsilon$ for some $\epsilon > 0$, and sells to consumer $\phi(\psi)$ at an assumed market-clearing price $p_{q'}$.\footnote{Note that as we showed $\Gamma$ is weakly increasing in $\psi$, the deviation can not change the ordering of firms in terms of the certificate they purchase and hence does not change the consumer to which they sell.} Then, the producer $\psi$ could instead deviate to purchasing $q_i$ at a price of $t_i$, and offering it on the competitive market at a price of $p_{q'} + g(q_i; \psi) - g(q'; \psi) + (t_i - t') + \epsilon/2$.  Note that if consumer $\phi(\psi)$ were to purchase from producer $\psi$ at this price, then her utility would be
\begin{align*}
& f(q_i; \phi(\psi)) - [p_{q'} + g(q_i; \phi(\psi)) - g(q'; \phi(\psi)) + (t_i - t') + \epsilon/2] \\
& = (f(q_i; \phi(\psi)) - g(q_i; \psi) - t_i - \epsilon) + (g(q'; \psi) + t') + \epsilon/2 - p_{q'} \\
& > f(q'; \phi(\psi)) - (g(q'; \psi) + t') + (g(q'; \psi) + t') - p_{q'} + \epsilon/2 \\
& > f(q'; \phi(\psi)) - p_{q'}.
\end{align*}
But since $(q', p_{q'})$ is the most-demanded offering to consumer $\phi(\psi)$ in the market equilibrium, this means that the offering of $q_i$ at the proposed price would be the most-demanded offering to consumer $\phi(\psi)$ under this deviation.  In particular this means that \emph{some} consumer would want to purchase $q_i$ at the suggested price, and therefore in the adjusted market equilibrium after this proposed deviation the price of $q_i$ must be at least this high. 

We conclude that the utility of producer $\psi$ under this deviation is at least
\begin{align*}
    [p_{q'} + (g(q_i; \psi) - g(q'; \psi) + (t_i - t') + \epsilon/2] - g(q_i; \psi) - t_i
    & = p_{q'} - g(q'; \psi) - t' + \epsilon / 2\\
    & > p_{q'} - g(q'; \psi) - t'
\end{align*}
and hence this deviation is strictly utility-improving for the producer, contradicting the assumption that $\Gamma$ is an equilibrium.


We therefore conclude that at equilibrium, each producer $\psi$ chooses whichever certificate $q_i$ from the menu maximizes $f(q_i; \phi(\psi)) - g(q_i; \psi) - t_i$.  The choice of each producer is therefore unique, up to tie-breaking on sets of measure zero.
\end{proof}

An immediate corollary of Lemma~\ref{lem:market.monotone.consumer} and Lemma~\ref{lem:market.monotone.producer} is that 
the equilibrium outcome for a given menu $M$ is not only essentially unique (up to the choice of market-clearing prices), but also has a natural assortative interpretation.  Each producer in the market has a corresponding consumer with whom they will always trade.  The producer will select whichever certification level maximizes the gains from trade between themselves and their partner consumer, less the price of the certification.

\begin{corollary}
\label{cor:market.assortative}
For any menu $M = \{(q_i, t_i)\}$ of the certifier, the resulting equilibrium market outcome has producer $\psi$ trade with consumer $\phi = \phi(\psi)$ where $G(\psi) = F(\phi)$.  The level of quality at which $\phi$ and $\psi$ trade maximizes $f(q_i; \phi) - g(q_i; \psi) - t_i$, and is weakly increasing in $\psi$.
\end{corollary}


%% file: revenue.tex
\section{Revenue-Optimal Certification}\label{sec:rev}

In the previous section we solved for the equilibrium outcome of the game played between producers and consumers given a certification menu $M$ chosen by the certifier.  With that characterization in hand, we now turn to the problem faced by a certifier who wishes to construct a revenue-maximizing slate of certificates.

\emph{A priori}, it would appear that the choice of which certificates to offer influences the downstream competitive market allocation and prices{. After all, producers compete for consumers and the certificates they purchase determine how they fare in this competition.} 
{Hence one might expect the menu of available certificates to influence} how much each producer would be willing to pay for any given certificate.  However, as we will show, the assortative characterization of market outcomes means that the certifier can reason about the behavior of each producer type separately, without worrying about how they will jointly interact in the competitive market. {Thus, the certifier is essentially facing a single buyer with an unknown type, and must simply maximize revenue subject to this buyer.  This leads us to define a reduction from the certifier's problem to that of a seller facing a (non-linear) buyer.}

\subsection{Reduction to a Non-linear Pricing Problem}

We will relate the certifier's revenue-maximization problem to a non-linear pricing problem between a single seller and a single buyer.  The buyer seeks to buy a perfectly divisible good. The seller may commit to a menu of quantities and prices. If a buyer {of type $\theta$} purchases quantity $q \in [0,1]$ of the good at a total price of $t$, then the buyer utility is
\begin{align*}
u((q,t); \theta) &= v(q; \theta) - t,
\end{align*}
where $v$ is concave {in quantity $q$} but not necessarily non-decreasing, and $v(0;\theta) = 0$ for all $\theta$.  The valuations $v$ satisfy a single-crossing condition, which is that for $\theta_1 < \theta_2$ and $q_1 < q_2$, we have 
\[ v(q_2; \theta_2) - v(q_1; \theta_2) > v(q_2; \theta_1) - v(q_1; \theta_1). \]
The principal seeks to maximize revenue subject to a constant cost of production $c > 0$ for any non-zero quantity {given a prior over buyer types}.  Given a menu $M$, we will write $\Rev(M)$ to denote the expected revenue obtained from $M$.  We will also write $\OPT$ for the revenue of the revenue-maximizing choice of menu $M$.

\begin{proposition}
\label{prop:reduction}
The certifier's revenue-maximization problem is equivalent to an instance of the non-linear pricing problem described above.
\end{proposition}
\begin{proof}
The certifier's problem is to design a certificate menu $M = \{(q_i, t_i)\}$ that maximizes the total revenue collected, less the certification cost $c$ paid to verify any non-trivial certificate $q_i > 0$.  By Corollary~\ref{cor:market.assortative}, given menu $M$, each producer $\psi$ will purchase whichever certificate $q_i$ maximizes $f(q_i; \phi(\psi)) - g(q_i; \psi) - t_i$.

For $q \in [0,1]$, we can interpret $\psi$ as a buyer type and define valuation function $v(q; \psi) = f(q_i; \phi(\psi)) - g(q_i; \psi)$.  Then since $f$ and $g$ both satisfy single-crossing with respect to their corresponding types, valuation function $v$ does as well.  Moreover, $v$ is concave and $v(0;\psi) = 0$.  By definition, the producers' choices of certificates from menu $M$ corresponds precisely to the buyer's choice of quantity when facing the same menu, interpreting each certificate quality threshold as a quantity.  Thus the outcomes, and hence revenue, in the two settings are equivalent.
\end{proof}

Note that an immediate implication of Proposition~\ref{prop:reduction}, given Corollary~\ref{cor:market.assortative}, is that for any menu $M$ chosen by the seller, the choice of quantity purchased by the buyer is weakly increasing in buyer type.\footnote{Alternatively, this is a direct consequence of the single-crossing condition on valuation functions $v$.}

\subsection{Optimizing Revenue in the Non-linear Pricing Problem}

By Proposition~\ref{prop:reduction}, to solve the certifier's revenue maximization problem it suffices to optimize revenue in the non-linear pricing problem.  As a first step, we note that in the special case where the valuations $v$ are linear in quantity (which happens, for example, if the cost $g$ and value $f$ functions in our certification problem are both linear in $q$), this problem reduces to a standard pricing problem in mechanism design.  A characterization due to Myerson~\cite{myerson1981optimal} then immediately establishes that it is revenue-optimal to choose a menu with only a single non-trivial item $(q,p)$ with $q = 1$. 

\begin{observation}
\label{obs:linear}
If $v(q;\theta)$ is linear in $q$ for all $\theta$, then there is a revenue-optimal menu that offers only quantity $q=1$ at some price $p$.  This price $p$ will be chosen to maximize $p \times \Pr_\theta[v(1;\theta) > p]$.
\end{observation}

However, in general, non-linearity substantially changes the problem relative to the linear case.  In particular, it is not necessarily optimal to offer a single menu item. 
\begin{proposition}
\label{prop:finite.suboptimal}
There are problem instances in which posting any single menu item is an arbitrarily poor approximation to the optimal revenue.  The approximation factor can be as large as $\Omega(\log(H))$, where $H = \max_{\theta_1, \theta_2}\frac{\max_q v(q; \theta_1)}{\max_q v(q; \theta_2)}$ is the ratio between the highest and lowest maximum values across buyer types.
\end{proposition}
\begin{proof}
Choose $c = 0$ and consider the valuation function $v(q;\theta) = q$ if $q \leq \theta$, and $v(q;\theta) = 2\theta - q$ if $q \geq \theta$.  That is, $v$ is piecewise linear for each $\theta$, with maximum value $\theta$ occurring at $q = \theta$.  Fix some $H \geq 1$ and suppose the type distribution is such that $\Pr[\theta > h] = 1/h$ for all $h \in [1,H]$.  That is, the type distribution is equal-revenue on range $[1,H]$.

This valuation function is concave and satisfies $v(0;\theta) = 0$ for all $\theta$.  Moreover, it satisfies the single-crossing condition.  Indeed, for any $q$ and any $\theta_1 < \theta_2$, we note that $\frac{d}{dq}v(q; \theta_1) \leq \frac{d}{dq}v(q; \theta_2)$, since for any $\theta$ this derivative is $1$ for $q < \theta$ and $-1$ for $q > \theta$.  Since $v$ is also continuous in both $q$ and $\theta$, the single-crossing condition is implied.\footnote{Technically our construction only satisfies weak single-crossing since the inequality in derivatives is not strict. We can make the example strict by perturbing the slope of the initial line segment by an infinitesimal amount so that it depends on the type $\theta$.  We omit these details for expositional clarity.}  Finally, we note that this valuation function $v$ can indeed arise in our reduction from the certification problem with producers and consumers.\footnote{In particular, take $\phi$ and $\psi$ to be supported on $[1,H]$, define $f(q; \phi) = q$ for all $\phi$ and $g(q; \psi) = \max\{0, 2(q-\psi)\}$ for all $\psi$.  Then $f$ is concave (in fact, linear), $g$ is convex, the single-crossing conditions are satisfied, and $v(q;\theta) = f(q; \phi(\theta)) - g(q; \theta)$ as required.}

We now show the desired gap in approximation.  Consider any menu $M$ with a single non-trivial menu item $(q,p)$.  Then the revenue achieved by the seller is at most the welfare generated by the optimal allocation of quality level $q$.  Since $v(q;\theta) \leq q$ for all $\theta$, this is certainly at most $q$ times the probability that $v(q;\theta) > 0$, which is $q \Pr[\theta > q/2] \leq q(2/q) = 2$.

On the other hand, the seller could offer a menu that includes every quality level $q \in [1,H]$ at a price of $q/2$.  A buyer of type $\theta$ would then choose to purchase quality level $q=\theta$ for a utility of $\theta/2$, generating revenue $\theta/2$.\footnote{Buying a higher quality level $q' > \theta$ is worse for the buyer because it generates less value at a higher price, whereas buying any quality level $q' <\theta$ generates utility $q' - q'/2 = q'/2 < \theta/2$.}  The total revenue is then $E[\theta/2] = O(\log H)$. 

Posting a single menu item is therefore at best an $O(\log H)$ approximation to the optimal revenue, as claimed.
\end{proof}

We therefore know that any approximately revenue-optimal mechanism must sometimes have multiple non-trivial offerings on its menu.  What should such a menu look like?  Note that since buyers do not have free disposal in our setting (i.e., valuations are non-monotone), it is not even immediately obvious that higher quantities should be sold for higher prices in a revenue-optimal menu.  We next establish that, in fact, there is always a revenue-optimal choice of menu for which higher quantities are sold at higher prices.

\begin{definition}
We say a menu $M = \{(q_i, p_i)\}$ is \emph{monotone} if for all $(q_i,p_i)$, $(q_j,p_j) \in M$ such that $q_i < q_j$, we have $p_i \leq p_j$.
\end{definition}


\begin{lemma}
\label{lem:monotone.prices}
For any instance of the non-linear pricing problem there is a revenue-optimal menu that is monotone.
\end{lemma}
\begin{proof}
Let $M$ be a revenue-optimal menu, and write $M = \{(q_i, p_i)\}_{i \in \Lambda}$ where $\Lambda$ is some (possibly uncountable) index set.  It is without loss to assume $\Lambda$ is a subset of $[0,1]$ such that $q_i < q_j$ for all $i < j$ (e.g., by reindexing so that the index of $q_i$ is equal to $q_i$).  We can further assume without loss of generality that every item in $M$ is purchased by some buyer type, as any element that is never purchased could be removed from $M$ without impact. 
By Corollary~\ref{cor:market.assortative}, quality levels purchased will be monotone non-decreasing in buyer type.  This means that every item $(q_i, p_i)$ is purchased by some contiguous interval of buyer types $I_i$ (which may have measure zero).


Suppose that menu $M$ is not monotone.  This means that either there is an element $i < \sup \Lambda$ such that $p_i > p_j$ for all $j > i$, or else there exists a pair of menu items $(q_i, p_i)$ and $(q_j, p_j)$ with $j > i$ such that (a) there exists some $\ell \in \Lambda$ with $i < \ell < j$, and (b) $p_\ell < \min\{p_i, p_j\}$ for all $i < \ell < j$.

Consider the former case, where there is an element $i < \sup \Lambda$ such that $p_i > p_j$ for all $j > i$.  In particular there must exist some $j \in \Lambda$ with $j > i$.
%
Let $M'$ be the menu $\{(q_\ell, p_\ell)\}_{\ell \in \Lambda, \ell \leq i}$.  I.e., $M'$ is $M$ with all elements with quantities greater than $q_i$ removed.  
Note that for all $j \leq i$, since the types $I_j$ preferred element $(q_j, p_j)$ to any other element in $M$, they prefer element $(q_j, p_j)$ to any other element in $M'$ as well.  Moreover, since purchase decisions are monotone in buyer type, we conclude that all types $\theta \in I_\ell$ with $\ell > i$ will purchase element $(q_i, p_i)$ from menu $M'$.  But since $p_i > p_\ell$ for all $\ell > i$, this means that the revenue generated by menu $M'$ is strictly greater than the revenue generated by menu $M$, contradicting the supposed optimality of menu $M$.

Next consider the other case, there exists a pair of menu items $(q_i, p_i)$ and $(q_j, p_j)$ such that $p_\ell < \min\{p_i, p_j\}$ for all $i < \ell < j$.  Let $M'$ be the menu $\{(q_\ell, p_\ell)\}_{\ell \in \Lambda, \ell \leq i} \cup \{(q_\ell, p_\ell)\}_{\ell \in \Lambda, \ell \geq j}$.  That is, $M'$ is menu $M$ with all elements 
strictly between $(q_i,p_i)$ and $(q_j,p_j)$ removed.  Then as in the previous case, for all $\ell \leq i$ and $\ell \geq j$, types $I_\ell$ all still prefer to purchase $(q_\ell,p_\ell)$.  In particular, types $I_i$ purchase $(q_i,p_i)$ and types $I_j$ purchase $(q_j,p_j)$.  By monotonicity of purchasing decisions due to Corollary~\ref{cor:market.assortative},
all intermediate types $\theta \in I_\ell$ for $i < \ell < j$ must purchase either $(q_i, p_i)$ or $(q_j, p_j)$.  As $p_i$ and $p_j$ are both larger than the prices of the elements those types were purchasing under menu $M$, the revenue of menu $M'$ must be strictly larger, which is again a contradiction.

We conclude that the prices in menu $M$ must be monotone non-decreasing in quality levels, as claimed.
\end{proof}

\subsection{An FPTAS for Revenue}

We are now ready to consider the problem of constructing an approximately revenue-optimal menu for the non-linear pricing problem.  For this we will make one further technical assumption, which is that the derivative of the valuations $v(q; \theta)$ with respect to $q$ is bounded at $0$.  I.e., there exists some $\lambda > 0$ such that $\frac{d}{dq}v(0; \theta) < \lambda$ for all $\theta$.  In other words, buyers are not infinitely sensitive to product quantity.  In the context of certification, this is implied by having $\frac{d}{dq}f(0; \phi) < \lambda$ for all $\phi$, meaning that consumers are not infinitely sensitive to quality at $0$.

\begin{assumption}
\label{ass:lipschitz}
There exists some $\lambda > 0$ such that $\frac{d}{dq}v(0; \theta) < \lambda$ for all $\theta$.
\end{assumption}

Given this assumption, the following result provides an FPTAS for the optimal menu with $k$ items.  We also show that by taking $k = 1/\epsilon$ one can obtain an FPTAS for the optimal (unrestricted) menu.

\begin{theorem}
\label{thm:fptas}
    A menu with revenue at least $\OPT - \lambda\epsilon$ can be found in time polynomial in $\lambda$ and $1/\epsilon$.  The menu can optionally be constrained to contain at most $k$ quality levels, in which case $OPT$ is the optimal revenue achievable with at most $k$ quality levels.
\end{theorem}


Our first step to proving Theorem~\ref{thm:fptas} is to show that we can restrict attention to menus with at most $1/\epsilon$ entries at only a small loss of revenue.  

\begin{lemma}
\label{lemma:finite}
For any monotone menu $M$, there exists a menu $M'$ of size at most $O(1/\epsilon)$ such that $\Rev(M') \geq \Rev(M) - O(\epsilon)$.
\end{lemma}
\begin{proof}
Fix the revenue-optimal menu $M = \{(q_i, p_i)\}_{i \in \Lambda}$.  
By Lemma~\ref{lem:monotone.prices} we can assume $M$ is monotone.  

We define $M'$ to be the following subset of $M$.  Let $A = \{\ell \epsilon\ \colon\ 0 \leq \ell \leq \lfloor 1/\epsilon\rfloor \}$.  Then for each $a \in A$, we add to $M'$ the element $(q,p)$ from $M$ with smallest $p$ such that $p \geq a$. Then we note that $M'$ contains at most $\lceil 1/\epsilon \rceil$ items.  Moreover, for each element $(q,p) \in M$ there is some $(q',p') \in M'$
such that $p' \geq p - \epsilon$.  

Any element in $M'$ will still be purchased by the types that purchased that element in $M$, as the menu of alternative options has only gotten smaller.  Since prices are monotone in quantity, and quantities purchased are monotone in buyer type, we can conclude that any type that was purchasing $(q,p)$ in the original menu $M$ will then purchase an element with price at least $p-\epsilon$ in menu $M'$.  The total loss in revenue conditional on any given buyer type is therefore at most $\epsilon$, and hence the total revenue loss is at most $\epsilon$ as well.
\end{proof}

Next, we show that we can discretize the possible (quality, price) pairs that appear in our menu without losing too much revenue. 

\begin{figure}
    \centering
\includegraphics[width=0.8\textwidth]{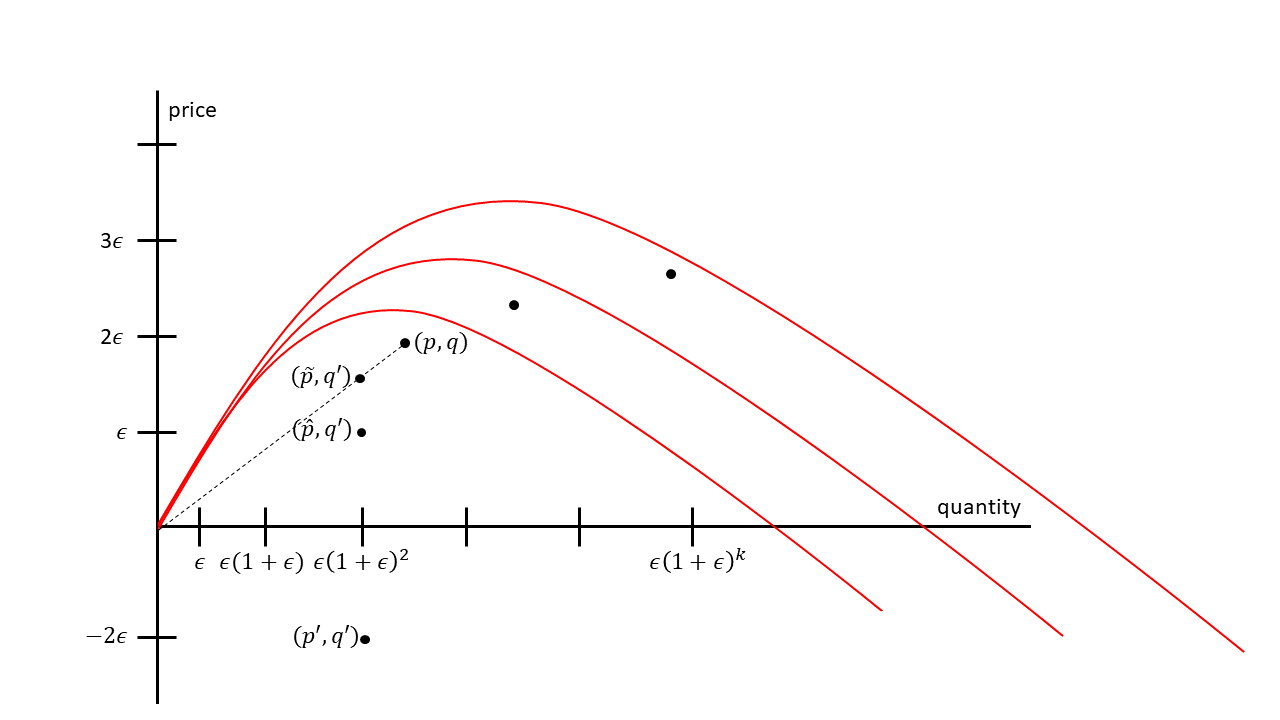}
    \caption{Illustration of the discretization procedure from Lemma~\ref{lemma:discretization}.  Red curves denote buyer valuation functions.  Menu item $(p,q)$ is discretized in quantities to $(\tilde{p},q')$ by shifting along a line to the origin, then discretized in price to $(\hat{p},q')$.  A final discount is applied to obtain the adjusted menu item $(p',q')$.}
    \label{fig:discrete}
\end{figure}

\begin{lemma}
\label{lemma:discretization}
For any monotone menu $M$ of size $k$, there exists a menu $M'$ such that
\begin{itemize}
    \item $M'$ has at most $k$ elements,
    \item for each $(q,p) \in M'$, $p$ is a multiple of $\epsilon$ and $q = \epsilon(1+\epsilon)^\ell$ for some integer $\ell \geq 0$, and
    \item $\Rev(M') \geq \Rev(M) - O((k+\lambda)\epsilon)$.
\end{itemize}
\end{lemma}
\begin{proof}
Let us first give the high-level idea for the construction, which is illustrated in Figure~\ref{fig:discrete}.  We will round each (quantity, price) pair on menu $M$ in three steps.  First, we will round each quantity down to an appropriate discretized grid, and then also lower the corresponding price \emph{to keep constant the ratio of quantity to price}.  The concavity of the value functions will then imply that buyer utilities cannot be reduced by too much, multiplicatively, as a result of this change.  This first step discretized the quantities; in the second step we discretize prices by rounding each price down to an appropriate grid, which can only increase utilities.  At this point we would be almost done, except for one complication: we must make sure that the (small) changes in buyer utility we induce do not result in buyers switching from more expensive items to significantly cheaper items from the menu.  It is here where we use the price-monotonicity of the menu.  Since the higher-quanitity items are the more expensive ones, in our third step we will provide discounts for the higher-quantity menu items, which will offset any utility perturbations due to discretization.  These discounts are what lead to the loss term being proportional to $k\epsilon$, rather than $\epsilon$, in our error bound.

We now move on to the formal construction.  Fix menu $M$ of size $k$, so that $M = \{(q_1, p_1), \dotsc, (q_k,p_k)\}$ where $q_1 < q_2 < \dotsc < q_k$.  We can assume without loss of generality that $p_1 \leq p_2 \leq \dotsc \leq p_k$, and that each element of $M$ is purchased with positive probability. 

We construct a new menu $M'$ in the following sequence of steps.  First, for each $(q_i, p_i)$ with $q_i \geq \epsilon$, let $q'_i$ be $q_i$ rounded down to the nearest value of $\epsilon(1+\epsilon)^\ell$ where $\ell \geq 0$ is an integer.  Then define $\tilde{p}_i = p_i \times (q'_i / q_i)$, noting that we chose $\tilde{p}_i$ so that $\tilde{p}_i/q'_i = p_i / q_i$.   This element $(q'_i, \tilde{p}_i)$ corresponds to rounding quantity but keeping the price to quantity ratio constant in our intuitive description above.  For our second step, we take $\hat{p}_i$ to be $\tilde{p}_i$ rounded down to the nearest multiple of $\epsilon$.  Finally, in our third step, we introduce our discounts.  For this we define $p'_i = \hat{p}_i - 3i\epsilon$, noting that the difference between $p'_i$ and $\hat{p}_i$ is increasing in $i$.  This completes our discretization procedure, so we add $(q'_i, p'_i)$ to menu $M'$.  We note that menu $M'$ is not necessarily monotone, may contain elements that are preferred by no types, and may contain elements with negative prices.

We claim that for every type $\theta$, if $\theta$ purchased item $(q_i, p_i)$ in menu $M$ with $q_i \geq \epsilon$, then $\theta$ will purchase some $(q'_j, p'_j)$ in menu $M'$ such that $j \geq i$.  To see why, first consider what would happen if $(q'_i, \tilde{p}_i)$ were on the menu.  What is $u((q'_i, \tilde{p}_i); \theta)$?  Since valuation function $v$ is concave and $v(0;\theta) = 0$, we must have $v(q'_i;\theta) \geq \frac{q'_i}{q_i}v(q_i;\theta)$.  But since $q'_i \geq \frac{1}{1+\epsilon}q_i$, this implies
\begin{align*}
    u((q'_i, \tilde{p}_i); \theta) &= v(q'_i; \theta) - \tilde{p}_i\\
    & \geq \frac{q'_i}{q_i}v(q_i;\theta) - \tilde{p}_i\\
    & = \frac{q'_i}{q_i}\left( v(q_i;\theta) - p_i \right)\\
    & \geq \frac{1}{1+\epsilon} u((q_i, p_i); \theta)\\
    & \geq u((q_i, p_i); \theta) - \epsilon.
\end{align*}
Since we also know that $\tilde{p}_i - \epsilon \leq \hat{p}_i \leq \tilde{p}_i$ and $p'_i = \hat{p}_i - 3i\epsilon$, we have
\begin{align*}
u((q'_i, p'_i); \theta) &= u((q'_i, \hat{p}_i); \theta) + 3i\epsilon\\
&\geq u((q'_i, \tilde{p}_i); \theta) + 3i\epsilon\\
&\geq u((q_i, p_i); \theta) + (3i-1)\epsilon.
\end{align*}
On the other hand, for any $j < i$, the utility of purchasing $(q'_j, \tilde{p}_j)$ is at most the utility of purchasing $(q_j, \tilde{p}_j)$, which is at most $\epsilon$ more than the utility of purchasing $(q_j, p_j)$ (since the ratio between $p_j$ and $\tilde{p}_j$ is no greater than $(1+\epsilon)$).  We therefore have
\begin{align*}
u((q'_j, p'_j); \theta) &= u((q'_j, \hat{p}_j); \theta) + 3j\epsilon\\
&\leq u((q'_j, \tilde{p}_j); \theta) + (3j+1)\epsilon\\
&\leq u((q_i, p_i); \theta) + (3j+2)\epsilon.
\end{align*}
Since we know that $u((q_i,p_i);\theta) \geq u((q_j,p_j);\theta)$ by assumption that $\theta$ purchases item $(q_i,p_i)$ from menu $M$, we conclude that $u((q'_i,p'_i);\theta) \geq u((q'_j,p'_j);\theta)$ as well, since $(3j+2) \leq (3i-1)$ for $i > j$. 

We conclude that each type $\theta$ that purchases $(q_i, p_i)$ from $M$ with $q_i \geq \epsilon$ will purchase a menu item $(q'_j, p'_j)$ from $M'$ such that $j \geq i$.  Since prices are monotone in menu $M$, we conclude that the total loss in revenue can be at most the difference in price between $p_i$ and $p'_i$ for any $i$.  This is at most $O(k\epsilon)$.

Finally, consider a type $\theta$ that purchases $(q_i, p_i)$ from $M$ with $q_i < \epsilon$.  By Assumption~\ref{ass:lipschitz}, the maximum willingness to pay for any agent for quality level $\epsilon$ is $\lambda \epsilon$.  These types therefore generate revenue at most  $\lambda \epsilon$, thus regardless of their purchase behavior they account for a total loss in revenue of at most $O(\lambda \epsilon)$.
\end{proof}

With Lemma~\ref{lemma:discretization} in hand, we can complete the proof of Theorem~\ref{thm:fptas} by employing dynamic programming to determine the revenue-optimal mechanism with a given maximum-quality entry.  One subtlety in the construction is that we must be careful to account for potential cannibalization by higher-quality elements in the menu.  We handle this by insisting that the menu we construct contains only elements that are selected by a non-zero measure of buyer types, and we check this condition when recursively applying the dynamic program.

\begin{proof}[Proof of Theorem~\ref{thm:fptas}]
We show how to compute the optimal menu with qualities and prices chosen from a discrete indexed set of possible options, using dynamic programming.  In general, given a quantity $q$ and price $p$ that lie in our discrete set of options, we will use $Q$ and $P$ to denote the integer indexing of $q$ and $p$, respectively.  

Given any choice of $Q$ and $P$ and some $k \geq 1$, write $M[Q,P,k]$ for the optimal revenue that can be obtained using a menu with at most $k$ elements, of which the one with highest quality is the one indexed by $Q$ and $P$.  We will also write $L[Q,P,k]$ for the lowest type $\theta$ that purchases quality level $Q$ in this optimal menu.  We can compute $M[Q,P,k]$ and $L[Q,P,k]$ recursively as follows.  If $k=1$ then $M[Q,P,k]$ is precisely $p$ times the probability that $v(q; \theta) \geq p$, and $L[Q,P,k]$ is precisely the infimum of types $\theta$ for which $v(q; \theta) \geq p$.

For $k > 1$, we will consider all possible options for the next-highest quality level on the menu given our discretization, say $(q',p')$ with $Q' < Q$.  For each choice of $(Q',P')$, we let $\theta(Q',P')$ denote the type that is indifferent between menu items $(Q',P')$ and $(Q,P)$, if any.  Recall from the single-crossing condition that this choice of $\theta(Q',P')$ is unique if it exists. 
If there is no such $\theta(Q',P')$, then we disqualify menu item $(Q',P')$ from consideration.  Otherwise, we consider $L[Q',P',k-1]$, the lowest type that purchases element $(Q',P')$ in the optimal menu with highest quality level $Q'$ at price $P'$.  If $L[Q',P',k-1] \geq \theta(Q',P')$, then again we disqualify menu item $(Q',P')$ from consideration, as this means that the optimal menu containing menu item $(Q,P)$ does not include any types that would purchase menu item $(Q',P')$.  

Otherwise, we have that $L[Q',P',k-1] < \theta(Q',P')$. We can therefore calculate the revenue from the optimal menu with highest and second-highest quality levels $(Q,P)$ and $(Q',P')$ as $R(Q',P') = M[Q',P',k-1] + (P-P')Pr[\theta > \theta(Q',P')]$.  That is, the additional revenue gain or loss due to including menu item $(Q,P)$ is $(P-P')Pr[\theta > \theta(Q',P')]$, the difference due to agents with type greater than $\theta(Q',P')$ switching from menu item $(Q',P')$ to menu item $(Q,P)$.

Finally, consider also the revenue that would be obtained by using only menu item $(Q,P)$; call this $R$.  If all potential choices of $(Q',P')$ were eliminated or if $R > R(Q',P')$ for all potential choices of $(Q',P')$, then we set $M[Q,P,k] = R$ and set $L[Q,P,k]$ to be the infimum type $\theta$ such that $v(Q; \theta) \geq P$.  This corresponds to the case that the optimal menu contains only the element $(P,Q)$.  Otherwise, let $(Q',P')$ be the choice that maximizes $R(Q',P')$, which by assumption is larger than $R$. Then we take $L[Q,P,k] = \theta(Q',P')$ and $M[Q,P,k] = R(Q',P')$.

We conclude that we can fill tables $M$ and $L$, with each entry taking time proportional to $\epsilon^{-2}$ (the time needed to consider every possible choice $(Q',P')$).  As there are $k\epsilon^{-2}$ entries in total, the total time to fill the tables is at most $k\epsilon^{-4}$.  The revenue-optimal mechanism with at most $k$ menu items can then be obtained by taking the maximum of $M[Q,P,n]$ over all choices of $Q$ and $P$.  

Finally, by Lemma~\ref{lemma:finite}, we can take $k = 1/\sqrt{\epsilon}$ and our dynamic program will obtain a menu $M$ such that $Rev(M)$ is at most $O(\lambda/\sqrt{\epsilon})$ less than the optimal revenue.  An appropriate change of variables, taking $k = 1/\epsilon$ and discretizing to multiples of $\epsilon^2$, then implies that our resulting menu is at most $O(\lambda\epsilon)$ less than that of the optimal menu.
\end{proof}

%% file: comparison.tex
\section{Welfare Maximization}\label{sec:comparison}

We have studied the problem faced by a revenue-maximizing certifier.  But what about a certifier that wishes to maximize the welfare enjoyed by the consumers and producers?  We could think of such a certifier as a government agency who is offering certification services not to generate profit, but rather to maximize the efficiency of production and trade.

We define the welfare of a market outcome as the sum of utilities of the consumers, the producers, and the certifier, taking into account all transfers between parties.  Given a menu $M$ of options provided by the certifier, we will write $\Wel(M)$ for the welfare that results in the unique market outcome resulting from menu $M$.

An immediate consequence of Corollary~\ref{cor:market.assortative} is that the welfare-optimal choice of menu is to offer all possible certification levels, at the cost of verification $c$.

\begin{theorem}
The welfare-optimal menu of certificates offers every possible certification level $q > 0$ at a cost of $c$, and level $0$ at a cost of $0$.
\end{theorem}
\begin{proof}
If all quality levels were visible, the welfare-maximizing outcome is would be for each producer $\psi$ to trade with consumer $\phi(\psi)$ at whichever quality level $q$ maximizes their gains from trade $f(q; \phi(\psi)) - g(q; \psi)$.  However, since quality levels are hidden, producers and consumers can trade at a positive level of quality only if the cost of verification is paid.  So if the maximum gains from trade is less than $c$, then it is preferable to trade at level $0$.  However, we observe that this is precisely the outcome implemented at equilibrium from the proposed certification menu, so it must be welfare-optimal over all possible menus.
\end{proof}

The welfare-optimal menu described above includes an arbitrarily large number of certificates.  In practice it may be helpful to find a welfare-optimal slate of at most $k$ certificates.  It turns out that a minor variation on the dynamic program described in the previous section can be used to compute an approximately welfare-optimizing slate, under Assumption~\ref{ass:lipschitz}.

\begin{theorem}
Let $M$ be the welfare-maximizing menu with at most $k$ elements.
A menu with $M'$ with at most $k$ elements and such that $\Wel(M') \geq \Wel(M) - O(\lambda\epsilon)$ can be found in time polynomial in $1/\epsilon$.
\end{theorem}
\begin{proof}
The proof is very similar to the one for Theorem~\ref{thm:fptas}, and strictly simpler, so we only briefly describe the differences here.  First, it is without loss to restrict attention to menus that post price $c$ for every non-trivial level of quality.  

Given any such menu $M$, we can discretize potential levels of quality by rounding down to the nearest multiple of $\epsilon$.  By Assumption~\ref{ass:lipschitz} this reduces welfare by at most $\lambda \epsilon$, as the welfare from each menu item is reduced by at most this much and each producer selects her gains-from-trade-maximizing element from the menu.

Given such a discretization, one can express the welfare-optimal menu recursively via dynamic programming as in Theorem~\ref{thm:fptas}, with the simplification that we need only index on quality rather than (quality, price) pairs (since all prices will be set to $c$).  Rather than defining $M[Q,k]$ (and respectively $L[Q,k]$) to be the maximum revenue of a menu with $k$ elements and maximum quality indexed by $Q$, it will be the maximum welfare of a menu with $k$ elements and maximum quality indexed by $Q$.  Our method of recursively computing $M[Q,k]$ (and $L[Q,k]$) then remains nearly unchanged relative to Theorem~\ref{thm:fptas}. The only change to note is the actual welfare calculation, relative to the revenue calculation.  In Theorem~\ref{thm:fptas} we used that the revenue obtained when agents of type $\theta > \theta'$ purchase certificate $q$ at price $p$ is $p$ times $\Pr[\theta > \theta']$.  In contrast, the welfare obtained when producers of type $\psi > \psi'$ all purchase certificate $q$ at price $c$ can be calculated in closed form as $\int_{\psi}^{\psi'} (f(q; \phi(\eta)) - g(q; \eta) - c)d\eta$.  Substituting this welfare calculation for the revenue calculation completes the necessary changes. 
\end{proof}

Is the revenue-maximizing choice of menu also approximately welfare-maximizing?  As it turns out, the welfare that results from the certifier's reveneu-optimal menu can be an arbitrarily small fraction of optimal welfare.  This is inherited from standard examples of monopolistic distortion in the linear settings, where a monopolist might be incentivized to sell much less of their good (i.e., certification) than what would be efficient in order to inflate prices.
\begin{proposition}
\label{prop:bad.welfare}
    There is a sequence of instances for the problem where welfare in the revenue-optimal solution is an arbitrarily small fraction of first-best welfare.  The approximation can be as bad as $\Omega(\log H)$, where $H = \max_{\theta_1, \theta_2}\frac{\max_q v(q; \theta_1)}{\max_q v(q; \theta_2)}$ is the ratio between the highest and lowest maximum values across buyer types.
\end{proposition}
\begin{proof}
Take $c = 0$ and suppose $f(q; \phi) = \phi q$ and $g(q; \psi) = 0$ for all $q$ and $\psi$.  Our distribution over consumer types $\phi$ is an equal-revenue distribution supported on $[1,H]$; that is, $F(\phi) = 1 - \frac{1}{\phi}$ for all $\phi \in [1,H]$.  Recalling Observation~\ref{obs:linear}, it is revenue-optimal for the certifier to offer contract $q=1$ at a price of $H$, for an expected welfare (and revenue) of $1$.  However, the optimal welfare, $\log(H)$, can be achieved by offering contract $q=1$ at price $0$.
%
%
%
%
\end{proof}

However, one implication of our equilibrium analysis is that adding additional certification options to a menu of certificates cannot reduce the sum of utilities of the consumers and producers, regardless of the prices selected.  One interpretation of this is that welfare cannot be harmed by a revenue-maximizing certifier entering a market for certification in which some certification options are already available.

\begin{proposition}
Consider two certification menus $M$ and $M'$ with $M \subseteq M'$.  Then $\Wel(M') - \Rev(M') \geq \Wel(M) - \Rev(M)$.
\end{proposition}
\begin{proof}
If producer $\psi$ selects option $(q,p)$ from certification menu $M$, then the welfare generated for the producer $\psi$ and corresponding consumer $\phi$, less the revenue raised by the certifier, is $f(q;\phi(\psi)) - g(q;\psi) - p$. By Corollary~\ref{cor:market.assortative}, producer $\psi$ purchases precisely whichever menu item from $M$ maximizes this quantity.  Providing additional items can therefore only increase the welfare jointly enjoyed by producer type $\psi$ and corresponding consumer $\phi(\psi)$.  As this holds pointwise for every $\psi$, it holds in aggregate over all types as well.
\end{proof}

